\documentclass{PoS}

\usepackage{amssymb,amsmath,amsthm,graphicx,subfigure}

\title{Properties of light pseudoscalars from lattice QCD with HISQ ensembles}

\ShortTitle{Properties of light pseudoscalars from HISQ ensembles}

\author{A. Bazavov and R.S. Van de Water\\
       Department of Physics, Brookhaven National Laboratory, Upton, NY 11973, USA\\
       \phantom{email}}

\author{C. Bernard and \speaker{M. Lightman}\\       
       Department of Physics, Washington University, St.~Louis, MO 63130, USA\\
       E-mail: \email{mlightman@physics.wustl.edu}\\}

\author{C. DeTar, L. Levkova and M. Oktay\\
       Department of Physics and Astronomy, University of Utah, Salt Lake City, UT 84112, USA\\
       \phantom{email}}

\author{W. Freeman, J. Kim and D. Toussaint\\
       Department of Physics, University of Arizona, Tucson, AZ 85721, USA\\
       \phantom{email}}

\author{Steven Gottlieb\\
        Department of Physics, Indiana University, Bloomington, IN 47405, USA\\
        and NCSA, University of Illinois, Urbana, IL 61801, USA\\
       \phantom{email}}

\author{U.M. Heller\\
        American Physical Society, One Research Road, Ridge, NY 11961, USA\\
       \phantom{email}}

\author{J.E. Hetrick\\
        Department of Physics, University of the Pacific, Stockton, CA 95211, USA\\
       \phantom{email}}

\author{J. Laiho\\
	SUPA, School of Physics \& Astronomy, University of Glasgow, Glasgow G12 8QQ, UK\\
       \phantom{email}}

\author{J. Osborn\\
        Argonne Leadership Computing Facility, Argonne National Laboratory, Argonne, IL 60439, USA\\
       \phantom{email}}

\author{R.L. Sugar\\
       Department of Physics, University of California, Santa  Barbara, CA 93106, USA\\
       \phantom{email}}

\vspace{-16mm}

\abstract{We fit lattice-QCD data for light-pseudoscalar masses and decay constants, from HISQ configurations generated by MILC, to SU(3) staggered chiral perturbation theory.  
At present such fits have rather high values of $\chi^2/{\rm d.o.f.}$,
possibly due to the lack of ensembles with 
lighter-than-physical sea strange-quark masses. We propose solutions to this problem for future work.  We also perform simple linear interpolations near the physical point on two ensembles with different lattice spacings, and obtain the preliminary result $(f_K/f_\pi )^{phys}=1.1872(41)$ in the continuum limit.}

\FullConference{XXIX International Symposium on Lattice Field Theory \\
                July 10 - 16 2011\\
                Squaw Valley, Lake Tahoe, California}

\begin{document}

\section{Introduction}
\vspace{-0.1 in}

A great deal of interesting physics is accessible from the detailed study of the
decay constants and masses of the light pseudoscalar mesons.  For example, a
determination of the decay constant ratio $f_K/f_\pi$ provides a competitive determination
of the CKM matrix element
$|V_{us}|$, but the lattice errors limit
the precision of the result \cite{PDG}. Similar fits can give ratios of light-quark
masses, and with nonperturbative or perturbative renormalization, absolute values of the
masses themselves. In addition, chiral fits make possible the determination 
of important low-energy constants.  Here we present a preliminary study of light 
pseudoscalar meson masses and decay constants using the highly improved staggered quark (HISQ) action \cite{original_HISQ}.  
The MILC collaboration has been generating HISQ configurations
\cite{HISQ_paper,HISQ_proceedings},
which are expected to improve upon results from the ``asqtad'' staggered quark action.  We attempt to fit the light pseudoscalar data with SU(3) staggered chiral perturbation theory (ChPT).  Since data exists for light- and strange-quark masses that are very nearly physical, we also perform simple linear interpolations of the data to the actual physical point and obtain a preliminary result for $(f_K/f_\pi)^{phys}$.

\vspace{-0.1 in}
\section{Ensembles and Correlators}
\vspace{-0.1 in}

Several ensembles have been generated by the MILC collaboration using the HISQ action, with 2+1+1 dynamical flavors and lattice spacings ranging from approximately 0.06 fm -- 0.15 fm.  The ensembles used in this report are found in Table \ref{tab:ensembles}.  We use a notation in which $m_l$ is the common mass of the up and down sea quark, $m_s$ the mass of the strange sea quark, and $m_c$ the mass of the charm sea quark.  Note that two ensembles have nearly physical $m_l$, $m_s$, and $m_c$.  Valence quarks are called $x$ and $y$, and their masses are denoted by $m_x$ and $m_y$ respectively.  Pseudoscalar mesons with valence quarks $x$ and $y$ have mass $m_{xy}$ and decay constant $f_{xy}$.  

Quark propagators were generated with various sources and sinks, with 8 -- 10 different valence-quark masses depending on the particular ensemble, and in the approximate range $m_l^{phys}$ -- $m_s^{phys}$, where $m_l^{phys}$ and $m_s^{phys}$ are the physical light- and strange-quark mass.  (Propagators with valence-quark mass $0.9m_c^{phys}$ and $m_c^{phys}$ were also generated for heavy-light physics, but were not used in this report.)
From these propagators, pseudoscalar meson correlators were calculated with random-wall and Coulomb-wall sources, and point sinks.  The random-wall and Coulomb-wall correlators were fit simultaneously with independent amplitudes and a common mass.  Only the amplitude from the random-wall correlators was used to obtain the decay constant.  Four different time slices on each lattice were used for the source locations, with the locations of these slices shifting as we went through successive configurations.

The fits started at a time $t_{min}$ from the source corresponding to $\sim 2.4$ fm in physical units for all of the lattice spacings.  This choice was made by examining, on all of the ensembles, the dependence of fitted quantities on $t_{min}$.  Fits including excited states and states of the opposite parity (``alternating states'') did not improve the quality of the fit, so they were not used.  The exception was one of the ensembles with a nearly physical light-quark mass, which required adding an alternating state.


\begin{table}[t]
\centering
\begin{tabular}{|c|c|c|c|c|}
\hline
$a$ (fm) & $am_l$ & $am_s$ & $am_c$ & $N_{lats}$ \\
\hline
0.15 & 0.013 & 0.065 & 0.838 & 1020 \\
0.15 & 0.0064 & 0.064 & 0.828 & 1000 \\
\hline
0.12 & 0.0102 & 0.0509 & 0.635 & 1040 \\
0.12 & 0.00507 & 0.0507 & 0.628 & 1020, 1000, 340* \\
0.12 & 0.00507 & 0.0304 & 0.628 & 1020 \\
0.12 & 0.00184 & 0.0507 & 0.628 & 500** \\
\hline
0.09 & 0.0074 & 0.037 & 0.44 & 1011 \\
0.09 & 0.00363 & 0.0363 & 0.43 & 633 \\
0.09 & 0.0012 & 0.0363 & 0.432 & 248** \\
\hline
0.06 & 0.0048 & 0.024 & 0.286 & 503 \\
0.06 & 0.0024 & 0.024 & 0.286 & 280 \\
\hline
\end{tabular}
\caption{HISQ ensembles used in this report.  A single star indicates three different ensembles that have all the same parameters but different spatial volumes ($L/a=24,32,40$).  Two stars indicate ensembles where all quark masses are nearly physical.  Note that all strange and charm quark masses are tuned to be nearly physical, except for the third ensemble with $a=0.12$ fm, 
where the strange-quark mass is taken to be approximately $3/5$ the physical value.}\label{tab:ensembles}
\end{table}

We attempt to account for autocorrelations in the covariance matrix of the fitted masses and decay constants by using the following \emph{scaled} covariance matrix in our chiral fits:
\begin{equation}
D_{ij}^{[b]}\equiv C_{ij}\sqrt{r_i^{[b]}r_j^{[b]}}, \quad\quad\quad \text{where} \quad r_i^{[b]}=\frac{C_{ii}^{[b]}}{C_{ii}}.
\end{equation}
Here $C_{ij}$ is an element of the original covariance matrix, and $C^{[b]}_{ij}$ is the corresponding element of the covariance matrix after blocking the configurations with block size $b$.  Note that the scaled covariance matrix only uses the diagonal elements of the blocked covariance matrix $C_{ii}^{[b]}$.  This is necessary because we use blocks of size $b=10$, and not all of the off-diagonal elements of $C_{ij}^{[b]}$ can be determined with good accuracy.

\section{Scale Setting}
\label{sec:Scale_Setting}

Before proceeding to do combined chiral fits of all the ensembles, it is necessary to determine accurately the \emph{relative} scale between ensembles with different lattice spacings.  
For each lattice spacing we take the ensemble with $m_l=0.1m_s$, and $m_s$ near the physical strange-quark mass.  
We look at masses $m_{xx}$, and decay constants $f_{xx}$, {\it i.e.}, for pseudoscalars with two valence quarks of equal mass $m_x$.  
Looking at the dimensionless ratio $f_{xx}/m_{xx}$ as a function of $m_x$, we find the value of $m_x$ for which $f_{xx}/m_{xx}=R$, where $R$ is some fiducial value.  
We take $R=0.3575$, which is the value of $f_{xx}/m_{xx}$ at $m_x=0.4m_s^{phys}$, $m_l=0.1m_s^{phys}$, and $m_s=m_s^{phys}$, as determined from a previous analysis using the asqtad action \cite{SU3_asqtad_proceedings}.  
We give to this value of $m_x$ the name ``$m_{p4s}$'', or just $m_p$ for short, and the values of $f_{xx}$ and $m_{xx}$ at this point are called ``$f_{pp}$'' and ``$m_{pp}$''.  
Thus we expect to find that $m_p\approx 0.4m_s$ on each of the HISQ ensembles we are using to set the relative scale, and the extent to which this is not true depends on the accuracy of the tuning of $m_s$ to the physical strange-quark mass, an issue to which we will return later.  (We ignore any possible mistuning of the charm mass as a small effect).

In practice, we determine $am_p$ by looking at the three sets of valence masses $am_x=0.3am_s$, $0.4am_s$, and $0.6am_s$, and putting a parabola through $m_{xx}^2/f_{xx}^2$ {\em vs.} $am_x$.  Once we have determined $am_p$ and $af_{pp}$ for a given lattice spacing, we can divide $af_{xy}$ and $am_{xy}$ by $af_{pp}$ for that lattice spacing, and similarly we can divide quark masses $am_x$, $am_y$, $am_l$, etc., by $am_p$.  This puts everything in ``p4s units''.  Quantities in p4s units are dimensionless quantities that are comparable between ensembles with different lattice spacings.%
\footnote{That is, if we knew $f_{pp}$ in MeV, then any meson mass or decay constant in p4s units from any lattice spacing would only need to be multiplied by this single number to put it in physical units.  A similar comment applies for $m_p$ and the quark masses.}
With all quantities in p4s units, a combined chiral fit can be carried out.  The advantage of our method is twofold.  First, since the unphysical ``p4s point'' is well within the parameters of the data set, it can be reached by simple interpolation.  Second, masses and decay constants at this point tend to have smaller errors than other quantities that could be used to set the scale.  Our method is similar to the scaling trajectory method of \cite{RBC_BK_continuum}.


\section{SU(3) Staggered Chiral Fits}

We use SU(3), partially quenched, rooted, staggered ChPT to fit the pseudoscalar masses and decay constants as a function of valence- and sea-quark mass.
The NLO expressions for the pseudoscalar mass and decay constant are given in Refs.~\cite{Aubin_Bernard_masses} and \cite{Aubin_Bernard_decay_const}, respectively.  In our fits, we also include NNLO continuum terms from Ref.~\cite{Bijnens_NNLO}, and add even higher-order analytic terms, which is similar to the previous asqtad analysis of Ref.~\cite{SU3_asqtad_proceedings}.  In order to obtain a reasonable fit to ChPT, we find it necessary to impose a cutoff on the valence quark masses: $m_x+m_y\le 0.6(m_l+m_s)$.  Also, our correlation matrix is close to singular, thus we smooth it by replacing all eigenvalues below $10^{-5}$ with their average.  This is done as a quick fix in order to obtain preliminary results, and in future fits we will thin the data set instead.

\begin{figure}[htp]
\centering
\subfigure[\label{sfig:decay_const}Pseudoscalar decay constant as a function of $m_x=m_y$.]{\includegraphics*[width=0.47\textwidth]{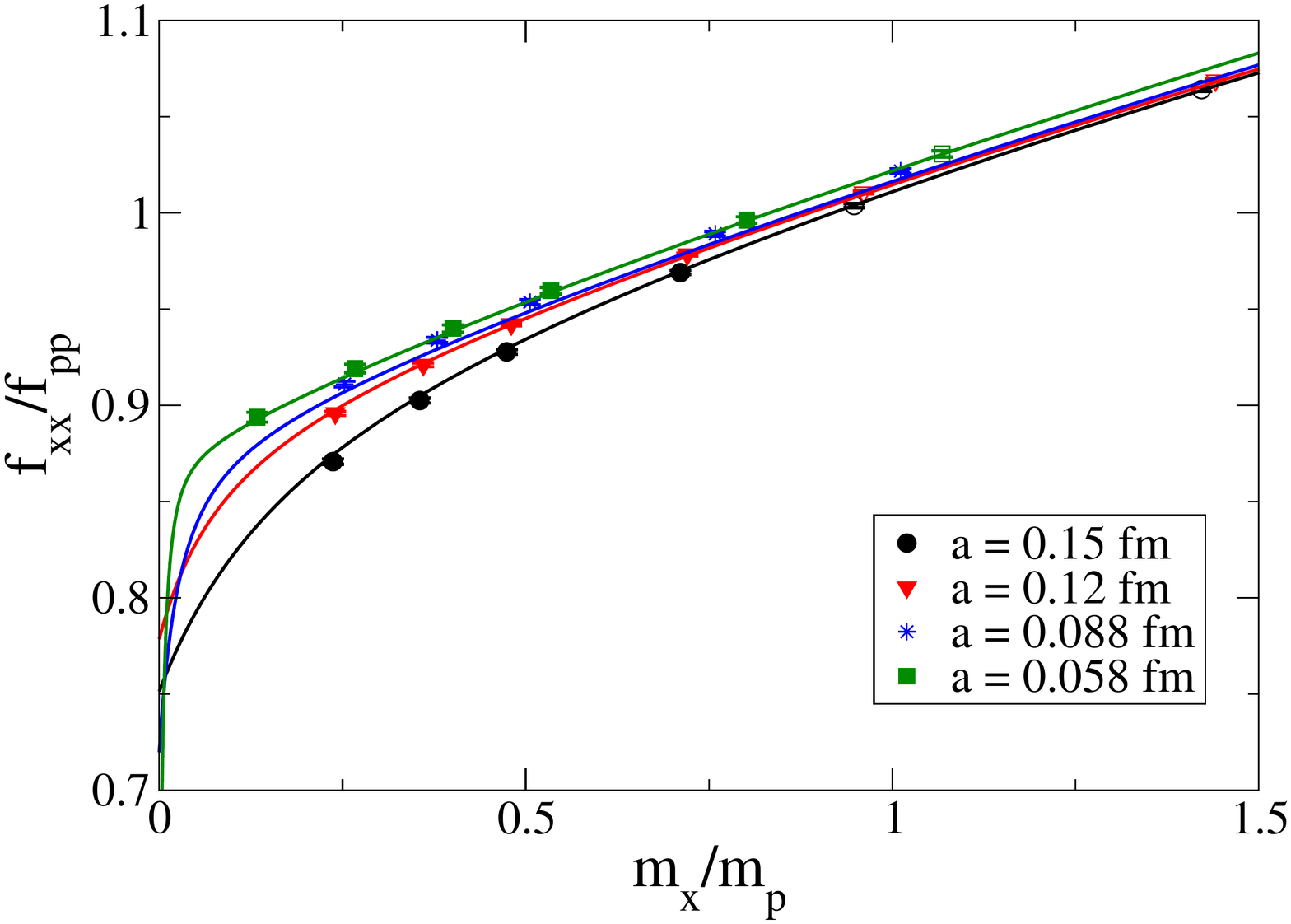}} \qquad
\subfigure[\label{sfig:mass}Pseudoscalar mass squared over the sum of the valence quark masses as a function of $m_x=m_y$.]{\includegraphics*[width=0.47\textwidth]{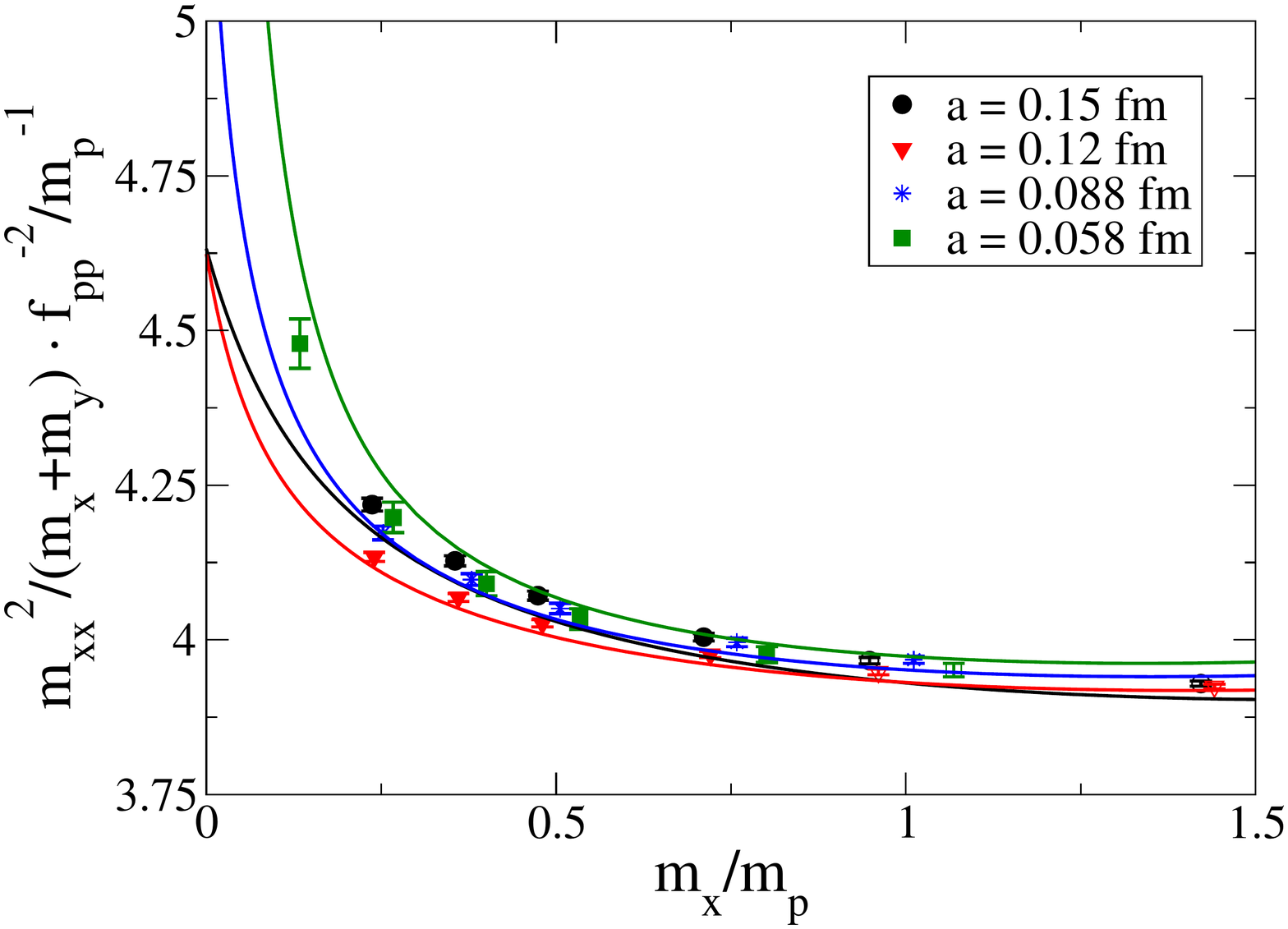}}
\caption{\label{fig:SU3_fit}Plots of the pseudoscalar mass and decay constant, and the fit to the data using SU(3) ChPT.  We show only pions in these plots, {\em i.e.} $m_x=m_y$ which is shown on the $x$-axis.  Data on the plots are for all of the different lattice spacings, (shown in different colors), but only for ensembles with $m_l=0.2m_s$.}
\end{figure}

We show the results of the fit on a subset of the data in Figure \ref{fig:SU3_fit}.  The fit seems quite good to the eye, however when the 
covariances are included we see that the fit is not satisfactory:  we get $\chi^2/\text{d.o.f.}=1747/380=4.6$.  If we do separate fits for each lattice spacing we get $\chi^2/\text{d.o.f.}=1.9$, 4.7, 2.6, and 0.7 for $a=0.15$ fm, 0.12 fm, 0.09 fm, and 0.06 fm respectively.  As discussed in Sec.~\ref{sec:Scale_Setting}, we assume that $m_s$ is equal to the physical strange-quark mass when we set the relative scale, which can be checked by seeing if $m_p/m_s=0.4$.  In fact, we find $0.37\lesssim m_p/m_s \lesssim 0.43$ for the different lattice spacings, indicating some mistuning of $m_s$.  However, this cannot completely explain the poor fit because the $\chi^2/\text{d.o.f.}$ is still large for fits to ensembles at single lattice spacings, where no scale setting is necessary.

Another possible reason for the lack of a good fit is that quark masses are too large for 
SU(3) ChPT to work well, at least at the order to which we are working.  In the asqtad analysis, this problem was dealt with by doing an initial fit with only lighter quark masses to fix LECs up to NNLO 
(and thereby controlling the extrapolation to $m_l^{phys}$), 
and then using the full set of quark masses to interpolate in the region
of $m_s^{phys}$ \cite{SU3_asqtad_proceedings}.  For the data set of this report (see Table \ref{tab:ensembles}),
we only have one lighter-than-physical value of $m_s$, so we cannot do something similar, at least in the sea sector.  New ensembles are being generated that will make this possible in future work.  Another possibility is to use SU(2) ChPT, in which the strange quark is considered heavy, to fit the data.  Staggered ChPT expressions for masses and decay constants have now been worked out for the SU(2) case~\cite{SU2_in_prep}.

\section{Linear Interpolations Near the Physical Point}
\label{sec:Linear_Interp}

\begin{figure}[htp]
\centering
\subfigure[\label{sfig:decay_const_linear}Plot of $f_{xy}$ vs. $m_x+m_y$.  Points corresponding to different fixed values of $am_y$ are indicated with different colors.]{\includegraphics*[width=0.47\textwidth]{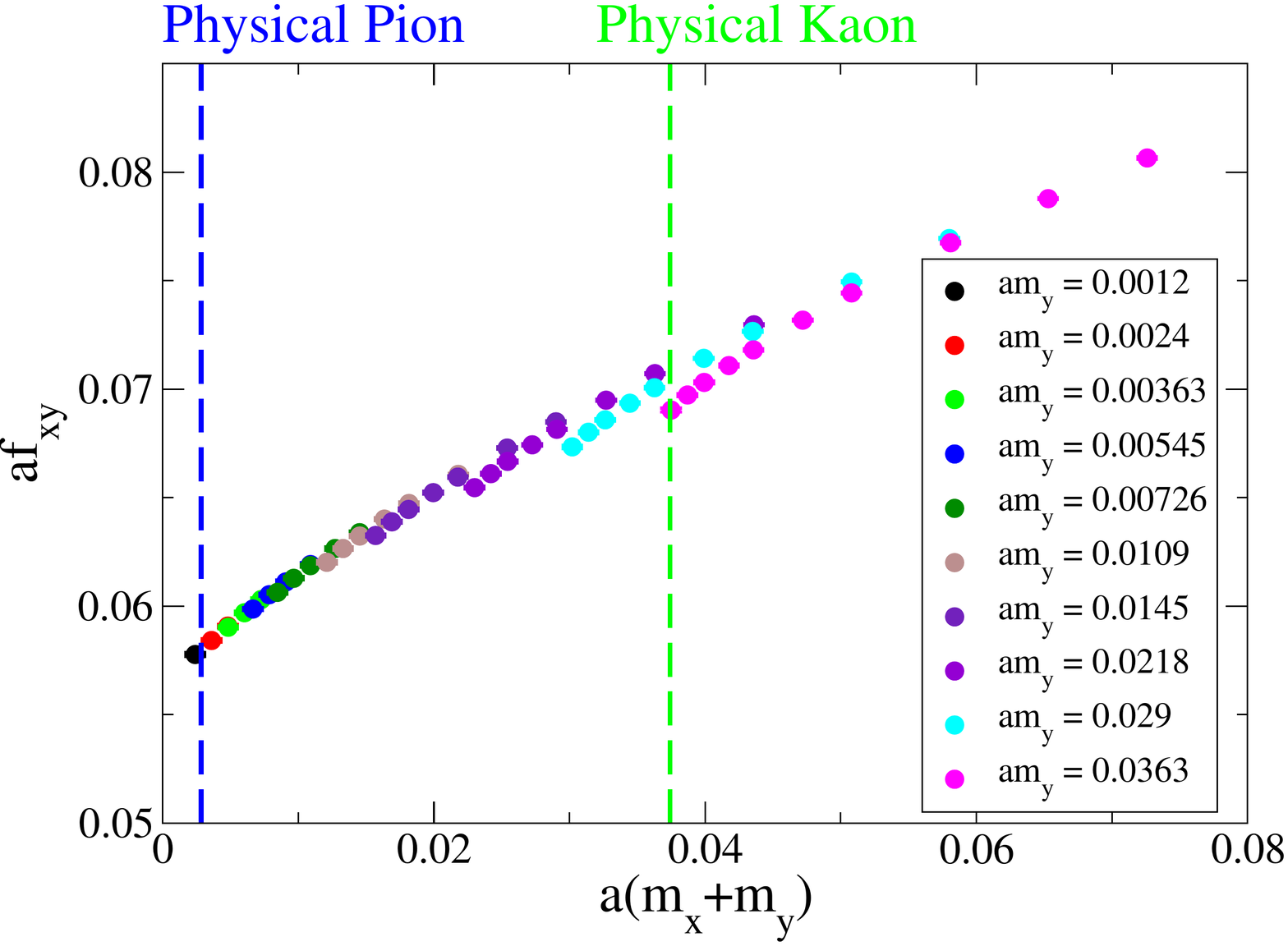}} \qquad
\subfigure[\label{sfig:mass_linear}Plot of $m_{xy}^2$ vs. $m_x+m_y$.  The unitary points for this ensemble are indicated in red.]{\includegraphics*[width=0.47\textwidth]{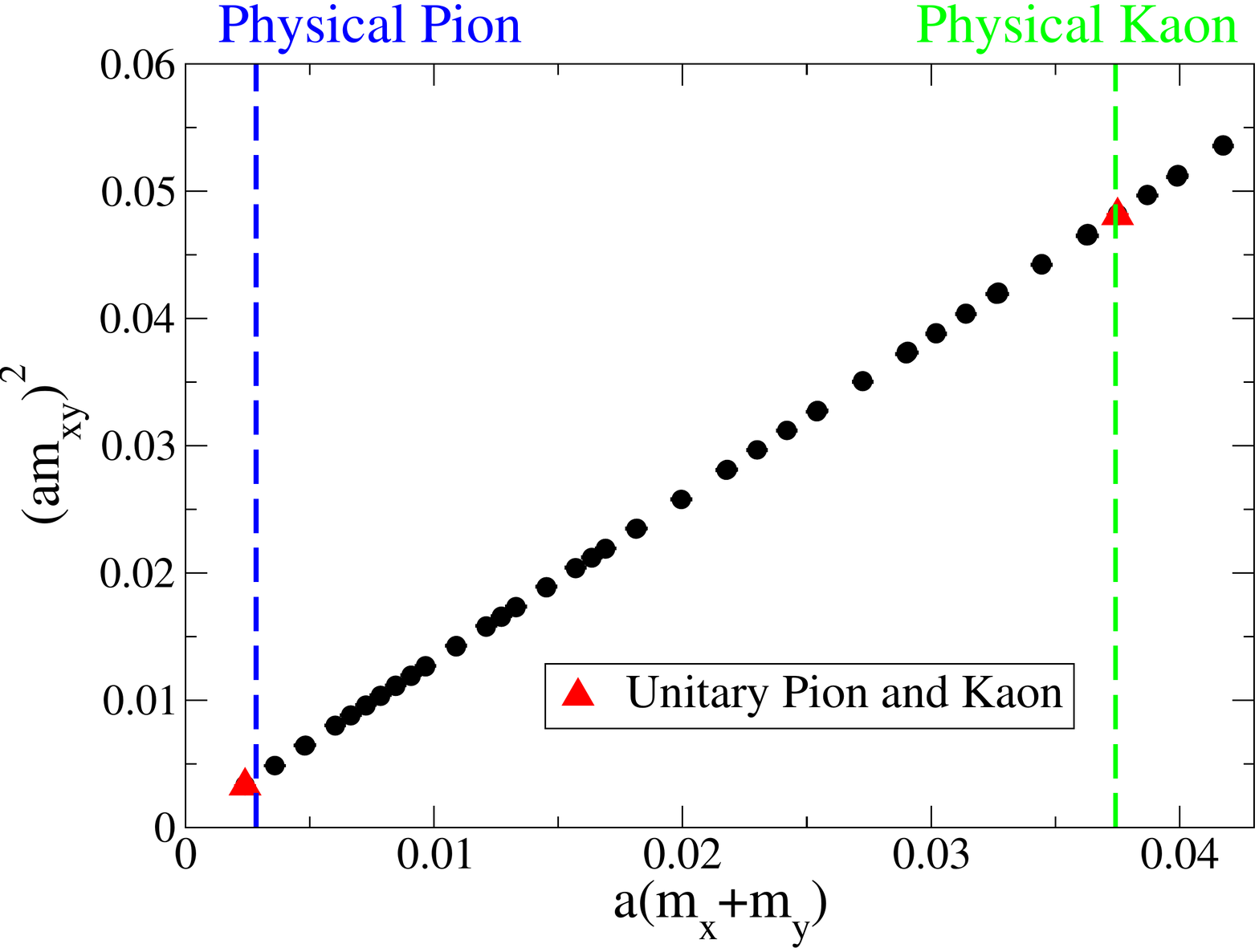}}
\caption{Plots of $m_{xy}^2$ and $f_{xy}$ vs. $m_x+m_y$ for the ensemble with $a=0.09$ fm, $am_l=0.0012$, $am_s=0.0363$, and $am_c=0.432$.  The values of $m_x+m_y$ corresponding to the physical pion and kaon mass are shown as blue and green dashed lines respectively.  These were determined using Eq.~\protect\ref{find_ml_ms_phys}.}\label{fig:linear_interp}
\end{figure}

In Table \ref{tab:ensembles} we see that there are two ensembles with a nearly physical light sea-quark mass.  Since we also have valence quarks with nearly physical light- and strange-quark masses, we can do a linear interpolation or extrapolation as necessary to the exact physical point.  Figure \ref{fig:linear_interp} shows plots of $m_{xy}^2$ and $f_{xy}$ vs. $m_x+m_y$ for one of the ensembles with nearly physical sea-quark masses.  The plot of $m_{xy}^2$ is remarkably linear.  The plot of $f_{xy}$ is linear for small quark masses, but when one of the valence-quark masses $m_y$ is heavier, the slope starts to depend on $m_y$.  The situation is similar for the other ensemble, not shown.  

The observations of the previous paragraph suggest the following analytic fit form for a single ensemble:
\begin{align}
(am_{xy})^2&=\left\{ \begin{array}{ll}
(am_\pi)^2(m_x,m_y)\equiv A_1+B_1a(m_x+m_y-2m_l), & (m_x,m_y)\text{ near }(m_l,m_l), \\
(am_K)^2(m_x,m_y)\equiv A_2+B_2a(m_x+m_y-m_l-m_s),  & (m_x,m_y)\text{ near }(m_l,m_s),
\end{array} \right. \label{Msq_linear} \\
af_{xy}&=\left\{ \begin{array}{ll}
af_\pi(m_x,m_y)\equiv C_1+D_1a(m_x+m_y-2m_l), & (m_x,m_y)\text{ near }(m_l,m_l), \\
af_K(m_x,m_y)\equiv C_2+D_2a(m_x-m_l)+E_2a(m_y-m_s), & (m_x,m_y)\text{ near }(m_l,m_s).
\end{array} \right. \label{F_linear}
\end{align}
We determine the values of the unknown coefficients $A_i$ -- $E_i$ using data for $am_{xy}$ and $af_{xy}$ at $(m_x,m_y)=(m_l,m_l)$, $(m_l, m_s)$, and the nearest neighboring points $(m_x,m_y)=(m_l,m^{(1)})$, $(m^{(1)},m_s)$, and $(m_l,m^{(2)})$, where $m^{(1)}$ is near $m_l$ and $m^{(2)}$ is near $m_s$.  For the ensemble with $a=0.09$ fm, we have $am_l=0.0012$, $am_s=0.0363$, $am^{(1)}=0.0024$, and $am^{(2)}=0.029$.

Keeping the sea-quark masses fixed by continuing to work within each ensemble, we then wish to determine $am_l^{phys}$ and $am_s^{phys}$, where $m_l^{phys}$ and $m_s^{phys}$ are the quark masses that correspond to the physical point.  Here we ignore the sea-quark masses entirely which introduces only a small partial quenching error because they are already close to physical.  We set
\begin{equation}\label{find_ml_ms_phys}
\frac{af_\pi(m_l^{phys},m_l^{phys})}{am_\pi(m_l^{phys},m_l^{phys})}=\frac{f_\pi^{exp}}{m_\pi^{exp}}, \quad\quad \frac{am_K(m_l^{phys},m_s^{phys})}{am_\pi(m_l^{phys},m_l^{phys})}=\frac{m_K^{exp}}{m_\pi^{exp}},
\end{equation}
where $f_\pi^{exp}$, $m_\pi^{exp}$, and $m_K^{exp}$ are the experimental values.  We can solve the left-hand equation for the unknown value $am_l^{phys}$, and then solve the right-hand equation to determine the unknown value $am_s^{phys}$.  This allows us to calculate the dimensionless ratio
\begin{equation}
\left(\frac{f_K}{f_\pi}\right)^{phys}=\frac{af_K(m_l^{phys},m_s^{phys})}{af_\pi(m_l^{phys},m_l^{phys})}.
\end{equation}

Results for $(f_K/f_\pi)^{phys}$ are shown in Table \ref{tab:FK_over_Fpi} for the two different ensembles with nearly physical sea-quark masses.  We also show a continuum extrapolation of $(f_K/f_\pi)^{phys}$ by assuming that it is linear in $a^2$ for non-zero lattice spacing.  To estimate the effect of partial quenching due to the fact that we kept sea-quark masses fixed during the analysis, we perform a similar analysis with unitary points only, from several of the ensembles with $a=0.12$ fm.%
\footnote{The reason this is not done in the first place is that it is only possible for $a=0.12$ fm with the current ensembles, (there is no lighter-than-physical strange-quark mass for the other lattice spacings), and because the interpolations are between points that are significantly further apart.}
We find $(f_K/f_\pi)^{phys}=1.1963(25)$, in agreement with the value of 1.1950(23) for the $a=0.12$ fm ensemble from Table \ref{tab:FK_over_Fpi}.  This gives us confidence that partial quenching is a small effect compared to the statistical uncertainty.

\begin{table}
\centering
\begin{tabular}{|c|c|c|c|c|c|}
\hline
$a$ (fm) & $am_l$ & $a\Delta m_l$ & $am_s$ & $a\Delta m_s$ & $(f_K/f_\pi)^{phys}$ \\
\hline
0.12 & 0.00184 & 0.00019 & 0.0507 & 0.0013 & 1.1950(23) \\
0.09 & 0.00120 & 0.00022 & 0.0363 & -0.0003 & 1.1913(16) \\
\hline
Continuum extrap. &  &  &  &  & 1.1872(41) \\
\hline
\end{tabular}
\caption{Results of interpolations to the physical point for $f_K/f_\pi$.  We perform a continuum extrapolation assuming that this quantity is linear in $a^2$ for non-zero lattice spacing.  We also show the distance of the interpolation points from the base data point, i.e. $\Delta m_l=m_l^{phys}-m_l$ and $\Delta m_s=m_s^{phys}-m_s$.}\label{tab:FK_over_Fpi}
\end{table}

\section{Conclusion}

We performed fits using SU(3) staggered ChPT to pseudoscalar masses and decay constants on HISQ lattices with several different lattice spacings.  We were unable to obtain a satisfactory fit, which could be due in part to strange-quark mistuning, and also possibly to the sea and valence strange-quark masses being too large for ChPT to work well.  New ensembles with lighter sea strange-quark masses will allow us to address these issues in future work by 1) allowing us to correct for strange-quark mistuning, and 
2) allowing us to use only lighter sea- and valence-quark masses to fix LECs to NNLO, and then to add higher order terms to interpolate near the strange mass using the heavier masses, which is an approach similar to Ref.~\cite{SU3_asqtad_proceedings}.  
We will also try using SU(2) staggered ChPT, which treats the strange quark as heavy.  We were able to obtain the preliminary result $(f_K/f_\pi)^{phys}=1.1872(41)$ using simple linear interpolations around data points with nearly physical quark masses.  The error on our result is comparable to that on the average of lattice results: $(f_K/f_\pi)^{phys}=1.193(5)$ \cite{FLAG, Jack_Ruth}, however we have included only the statistical error in our result.  
When the full fit to a much larger set of data points is performed, we expect to be able to obtain a rather precise result.

Computations for this work were carried
out with resources provided by the USQCD Collaboration, the ALCF, and NERSC, which are funded by the Office of Science of the U.S. Department of Energy; and with resources provided by the NCSA and NICS, SDSC, and TACC, which are funded through the National Science Foundation's Teragrid/XSEDE Program. This work was supported in part by the U.S Department of Energy and the U.S. National Science Foundation.

\vspace{-0.1 in}
\bibliography{Lightman_Lattice2011}

\end{document}